\documentclass[12pt]{article}
\usepackage{cite}
\usepackage{epsfig}

\catcode`@=11
\def\citer{\@ifnextchar [{\@tempswatrue\@citexr}{\@tempswafalse\@citexr[]}}
 
\def\@citexr[#1]#2{\if@filesw\immediate\write\@auxout{\string\citation{#2}}\fi
  \def\@citea{}\@cite{\@for\@citeb:=#2\do
    {\@citea\def\@citea{--\penalty\@m}\@ifundefined
       {b@\@citeb}{{\bf ?}\@warning
       {Citation `\@citeb' on page \thepage \space undefined}}%
\hbox{\csname b@\@citeb\endcsname}}}{#1}}
\catcode`@=12

\def\NPB#1#2#3{{\it Nucl.~Phys.} {\bf{B #1}} (19#2) #3}
\def\PLB#1#2#3{{\it Phys.~Lett.} {\bf{B #1}} (19#2) #3}
\def\PRD#1#2#3{{\it Phys.~Rev.} {\bf{D #1}} (19#2) #3}

\def\refeq#1{\mbox{eq.~(\ref{#1})}}

\def\citere#1{\mbox{Ref.~\cite{#1}}}
\def\citeres#1{\mbox{Refs.~\cite{#1}}}

\newcommand{\MstL}{M_{\tilde{t}_L}}
\newcommand{\MstR}{M_{\tilde{t}_R}}
\newcommand{\MsbL}{M_{\tilde{b}_L}}
\newcommand{\MsbR}{M_{\tilde{b}_R}}

\newcommand{\At}{A_t}
\newcommand{\Ab}{A_b}
\newcommand{\Xt}{X_t}
\newcommand{\Xb}{X_b}

\newcommand{\ms}{M_S}
\newcommand{\msusy}{M_{\mathrm{SUSY}}}

\newcommand{\msbar}{$\overline{\rm{MS}}$}
\newcommand{\MS}{\overline{\mathrm{MS}}}
\newcommand{\OS}{\mathrm{OS}}

\newcommand{\oas}{{\cal O}(\alpha_s)}

\newcommand{\cp}{{\cal CP}}

\newcommand{\edz}{\frac{1}{2}}

\newcommand{\twol}{two-loop}
\newcommand{\onel}{one-loop}

\newcommand{\fh}{{\em FeynHiggs}}
\newcommand{\fhtwo}{{\em FeynHiggs2.0}}

\newcommand{\subh}{{\em subhpole}}
\newcommand{\subhtwo}{{\em subhpole2}}

\newcommand{\MZ}{M_Z}
\newcommand{\MA}{M_A}
\newcommand{\mh}{m_h}
\newcommand{\mhmax}{m_h^{\rm max}}

\newcommand{\mt}{m_{t}}
\newcommand{\mtms}{\overline{m}_t}
\newcommand{\mtexp}{m_t^{\rm exp}}
\newcommand{\mtbar}{\overline{m}_{t}}

\newcommand{\mb}{m_{b}}

\newcommand{\mgl}{m_{\tilde{g}}}

\newcommand{\Stop}{\tilde{t}}
\newcommand{\StopL}{\tilde{t}_L}
\newcommand{\StopR}{\tilde{t}_R}

\newcommand{\Sbot}{\tilde{b}}
\newcommand{\SbotL}{\tilde{b}_L}
\newcommand{\SbotR}{\tilde{b}_R}

\newcommand{\tsf}{\theta\kern-.20em_{\tilde{f}}}
\newcommand{\tsfp}{\theta\kern-.20em_{\tilde{f}\prime}}
\newcommand{\tsq}{\theta\kern-.15em_{\tilde{q}}}
\newcommand{\sw}{s_W}
\newcommand{\cw}{c_W}

\newcommand{\VL}{\left( \begin{array}{c}}
\newcommand{\VR}{\end{array} \right)}
\newcommand{\ML}{\left( \begin{array}{cc}}
\newcommand{\MLd}{\left( \begin{array}{ccc}}
\newcommand{\MLv}{\left( \begin{array}{cccc}}
\newcommand{\MR}{\end{array} \right)}

\newcommand{\Tb}{\tan \beta\hspace{1mm}}
\newcommand{\tb}{\tan \beta}

\newcommand{\CTb}{\cot \beta\hspace{1mm}}

\newcommand{\CZb}{\cos 2\beta\hspace{1mm}}

\newcommand{\tev}{\,\, \mathrm{TeV}}
\newcommand{\gev}{\,\, \mathrm{GeV}}

\newcommand{\BC}{\begin{center}}
\newcommand{\EC}{\end{center}}
\newcommand{\BE}{\begin{equation}}
\newcommand{\EE}{\end{equation}}
\newcommand{\BEA}{\begin{eqnarray}}
\newcommand{\BEAnn}{\begin{eqnarray*}}
\newcommand{\EEA}{\end{eqnarray}}
\newcommand{\EEAnn}{\end{eqnarray*}}
\newcommand{\non}{\nonumber}
\newcommand{\id}{{\rm 1\kern-.12em
\rule{0.3pt}{1.5ex}\raisebox{0.0ex}{\rule{0.1em}{0.3pt}}}}
\newcommand{\lsim}
{\;\raisebox{-.3em}{$\stackrel{\displaystyle <}{\sim}$}\;}
\newcommand{\gsim}
{\;\raisebox{-.3em}{$\stackrel{\displaystyle >}{\sim}$}\;}

\def\als{\alpha_s}

\newcommand{\hbb}{h \to b\bar{b}}
\newcommand{\Hbb}{H \to b\bar{b}}

\begin{document}
\thispagestyle{empty}

\def\thefootnote{\fnsymbol{footnote}}

\begin{flushright}
CERN-TH/99-374\\
DESY 99-186\\
hep-ph/9912223
\end{flushright}

\vspace{1cm}

\begin{center}

{\large\sc {\bf Suggestions for Improved Benchmark Scenarios}}

\vspace*{0.4cm} 

{\large\sc {\bf for Higgs-Boson Searches at LEP2%
\footnote{Contribution to the Workshop on ``New Theoretical Developments
for Higgs Physics at LEP2'', CERN, Geneva, Switzerland, October 1999. \\[.2em]}
}}

\vspace{1cm}

{\sc 
M.~Carena$^{1}$%
\footnote{On leave from the Theoretical Physics Department,
Fermilab, Batavia, 
IL 60510-0500, USA} %
, S.~Heinemeyer$^{2}$%
, C.E.M.~Wagner$^{1}$%
\footnote{On leave from the High Energy Division, Argonne
National Laboratory, Argonne,  
IL 60439, USA\\
and the Enrico Fermi Institute, Univ.\ of Chicago, 5640 Ellis, Chicago,
IL 60637, USA} 
~and G.~Weiglein$^{1}$%
}

\vspace*{1cm}

{\sl
$^1$ Theoretical Physics Division, CERN, CH-1211 Geneva 23, Switzerland

\vspace*{0.4cm}

$^2$ DESY Theorie, Notkestr. 85, 22603 Hamburg, Germany }

\end{center}

\vspace*{2cm}

\begin{abstract}
We suggest new benchmark scenarios for the Higgs-boson search at LEP2.
Keeping $\mt$ and $\msusy$ fixed, we improve on the definition of
the maximal mixing benchmark scenario defining precisely the values of all 
MSSM parameters such that the new $\mhmax$
benchmark scenario yields the parameters which 
maximize the value of $\mh$ for a given $\tb$. %within the MSSM. 
The corresponding scenario with vanishing mixing in the scalar top sector is
also considered. We propose a further benchmark scenario with a relatively 
large value of
$|\mu|$, a moderate value of $\msusy$, and moderate mixing parameters 
in the scalar top sector. While the latter scenario yields $\mh$ values
that in principle allow to access the complete $\MA$--$\tb$-plane at
LEP2, on the other hand it contains parameter
%allows to access 
%parts of the parameter space in the $\MA$--$\tb$-plane with relatively 
%large values of $\tb$, it on the other hand also contains parameter 
regions where the Higgs-boson detection can be difficult, because of a
suppression of the branching ratio of its decay into bottom quarks.
%owing to a suppression of $BR(\hbb)$ the Higgs-boson
%detection can be difficult.
\end{abstract}
%\pacs{}

\vspace{.5cm}

December 1999\\[2em]

\def\thefootnote{\arabic{footnote}}
\setcounter{page}{0}
\setcounter{footnote}{0}

\newpage

%%%%%%%%%%%%%%%%%%%%%%%%%%%%%%%%%%%%%%%%%%%%%%%%%%%%%%%%%%%%%%
%%%%%%%%%%%%%%%%%%%%%%%%%%%%%%%%%%%%%%%%%%%%%%%%%%%%%%%%%%%%%%

\section{Introduction and theoretical basis}

Within the MSSM the masses of the $\cp$-even neutral Higgs bosons are
calculable in terms of the other MSSM parameters. The mass of the
lightest Higgs boson, $\mh$, has been of particular interest as it is
bounded from above according to $\mh \leq \MZ$ at the tree level.
The radiative corrections at \onel\ order~\cite{mhiggs1l,mhiggsf1l} have 
been supplemented in the last years with the leading \twol\ corrections, 
performed by renormalization group (RG) 
methods~\cite{LL,mhiggsRG1a}, by renormalization 
group improvement of the  one-loop effective potential 
calculation~\cite{mhiggsRG1b,mhiggsRG2}, 
by two-loop effective  potential calculations~\cite{hoanghempfling,zhang},   
and in the Feynman-diagrammatic (FD)
approach~\cite{mhiggsletter,mhiggslong}. These calculations predict an
upper bound for $\mh$ of about $\mh \lsim 130 \gev$.%
\footnote{
This value holds for $\mt = 175 \gev$ and $\msusy = 1 \tev$. If $\mt$
is raised by $5 \gev$ then the $\mh$ limit is increased by about $5 \gev$; 
using $\msusy = 2 \tev$ increases the limit by about $2 \gev$.
}

The numerical evaluations of the neutral $\cp$-even Higgs-boson
masses are implemented in two Fortran codes that 
are used for phenomenological studies by the LEP collaborations: the
program \subh, corresponding to the RG calculation~\cite{mhiggsRG1b}, 
and the program \fh~\cite{feynhiggs}, corresponding to the result of 
the FD calculation.

The tree-level value for $\mh$ within the MSSM is determined by $\tb$, 
the $\cp$-odd Higgs-boson mass $\MA$, and the $Z$-boson mass $\MZ$. 
Beyond the tree-level, the main correction to $\mh$ stems from the 
$t$--$\Stop$-sector, and for large values of $\tb$ also from the 
$b$--$\Sbot$-sector.

In order to fix our notations, we list the conventions for the inputs
from the scalar top and scalar bottom sector of the MSSM:
the mass matrices in the basis of the current eigenstates $\StopL, \StopR$ and
$\SbotL, \SbotR$ are given by
\BEA
\label{stopmassmatrix}
{\cal M}^2_{\Stop} &=&
  \ML \MstL^2 + \mt^2 + \CZb (\edz - \frac{2}{3} \sw^2) \MZ^2 &
      \mt \Xt \\
      \mt \Xt &
      \MstR^2 + \mt^2 + \frac{2}{3} \CZb \sw^2 \MZ^2 
  \MR, \\
&& \non \\
\label{sbotmassmatrix}
{\cal M}^2_{\Sbot} &=&
  \ML \MsbL^2 + \mb^2 + \CZb (-\edz + \frac{1}{3} \sw^2) \MZ^2 &
      \mb \Xb \\
      \mb \Xb &
      \MsbR^2 + \mb^2 - \frac{1}{3} \CZb \sw^2 \MZ^2 
  \MR,
\EEA
where 
\BE
\mt \Xt = \mt (\At - \mu \CTb) , \quad
\mb\, \Xb = \mb\, (\Ab - \mu \Tb) .
\label{eq:mtlr}
\EE
Here $A_t$ denotes the trilinear Higgs--stop coupling, $A_b$ denotes the
Higgs--sbottom coupling, and $\mu$ is the Higgs mixing parameter.

SU(2) gauge invariance leads to the relation
\BE
\MstL = \MsbL .
\EE
For the numerical evaluation, a convenient choice is
\BE
\MstL = \MsbL = \MstR = \MsbR =: \msusy ;
\label{eq:msusy}
\EE
this has been shown to yield upper values for $\mh$ which comprise
also the case where $\MstR \neq \MstL \neq \MsbR$,
when $\msusy$ is identified with the heaviest one~\cite{mhiggslong}.
We furthermore use the short-hand notation
\BE
\ms^2 := \msusy^2 + \mt^2~.
\EE

Accordingly, the most important parameters for the corrections to $\mh$
are $\mt$, $\msusy$, $\Xt$, and $\Xb$. The mass of the lightest
$\cp$-even Higgs boson depends furthermore on the SU(2) gaugino mass
parameter, $M_2$. The other gaugino mass parameter, $M_1$, is usually
fixed via the GUT relation 
\BE
M_1 = \frac{5}{3} \frac{\sw^2}{\cw^2} M_2.
\EE
At the two-loop level also the gluino mass, $\mgl$, enters the
prediction for $\mh$. In \fh\ the gluino mass can be specified as a free
input parameter. The effect of varying $\mgl$ on $\mh$ is up to $\pm 2 \gev$ 
for large mixing in the $\Stop$-sector and below $\pm 0.5 \gev$ for
vanishing mixing~\cite{mhiggslong}. Within  
\subh, the gluino mass was
fixed to $\mgl = \msusy$. Compared to the maximal values of $\mh$ 
(obtained for $\mgl \approx 0.8\,\msusy$) this leads to a reduction of the 
Higgs-boson mass by up to $0.5 \gev$. Within the new version, \subhtwo,
arbitrary values of the gluino mass will be allowed as input.

It should be noted in this context that the FD result has been obtained
in the on-shell (OS) renormalization scheme, whereas the RG result has been 
calculated using the \msbar\ scheme. Owing to the different
schemes used in the FD and the RG approach for the
renormalization in the scalar top sector, the
parameters $\Xt$ and $\msusy$ are also scheme-dependent
in the two approaches. This difference between the corresponding
parameters has to be taken into account 
when comparing the results of the two approaches. In a simple approximation
the relation between the parameters in the different schemes is given
by~\cite{bse}
\BEA
\ms^{2, \MS} &=& \ms^{2, \OS} 
 - \frac{8}{3} \frac{\alpha_s}{\pi} \ms^2 , 
\label{eq:msms} \\
\Xt^{\MS} &=& X_t^{\OS} + \frac{\alpha_s}{3 \pi} \ms 
   \left(8 + 4 \frac{X_t}{\ms} - 3 \frac{X_t}{\ms} 
   \ln\left(\frac{\mt^2}{\ms^2}\right) \right),
%X_t^{\MS} &=& X_t^{\OS} \frac{\mt^{\OS}}{\mt^{\MS}(\ms)} +
%  \frac{8}{3} \frac{\alpha_s}{\pi} \ms 
\label{eq:xtms} 
%\mtms^{\MS}(\mu) &=& 
%  \mtbar \left(1 + \frac{\alpha_s}{\pi} 
%    \ln\left(\frac{\mt^2}{\mu^2}\right) \right) ,
%     \label{eq:mtsmreg} 
\EEA 
where in the terms proportional to $\alpha_s$ it is not necessary to
distinguish between \msbar\ and on-shell quantities, since the
difference is of higher order.
The \msbar\ top-quark mass, $\mt^{\MS}(m_t) \equiv \mtms$, 
is related to the top-quark pole
mass, $\mt^{\OS} \equiv \mt$, in $\oas$ by
\BE
\mtms = \frac{\mt}{1 + \frac{4}{3\,\pi} \als(\mt)}~.
\label{mtrun}
\EE
While the resulting shift in the parameter $\msusy$ turns out to be 
relatively small in general, sizable differences can occur between the 
numerical values of 
$\Xt$ in the two schemes, see \citeres{mhiggslong,bse}. For this reason
we specify below different values for $\Xt$ within the two approaches.

The results of the RG and the FD calculation have been compared in detail in
\citeres{mhiggslle,bse}. While the results agree in the logarithmic
terms at the two-loop level~\cite{bse}, the FD result (program \fh)
contains further genuine two-loop corrections that are not present in the RG
calculation. These corrections lead to an increase in the maximal 
values for $\mh$ by up to $4 \gev$. Within the 
one-loop effective potential
computation, for large values of $\MA$ and $\ms$, $\mgl = \msusy$,
and not too large 
mixing parameters in the scalar top sector, the bulk of these 
corrections is taken 
into account by incorporating the proper one-loop relation between the
running top quark mass at the scale $\mt$ and the one at the scale
$\ms$ when computing the finite threshold corrections to the
Higgs quartic coupling at the scale $\ms$~\cite{bse}. The proper
relation between $\mtms$ and $\mt(\ms)$
will be used in the program based on  the RG improved one-loop
effective potential calculation (\subhtwo)
available for public use in the near future.

%%%%%%%%%%%%%%%%%%%%%%%%%%%%%%%%%%%%%%%%%%%%%%%%%%%%%%%%%%%%%%
%%%%%%%%%%%%%%%%%%%%%%%%%%%%%%%%%%%%%%%%%%%%%%%%%%%%%%%%%%%%%%

\section{The benchmark scenarios}

By combining the theoretical result for the upper bound on $\mh$ as a 
function of $\tb$ within the MSSM with the informations from the direct search
for the lightest Higgs boson, it is possible to derive constraints on
$\tb$. Since the predicted value of $\mh$ depends
sensitively on the precise numerical value of $\mt$, it has become
customary to discuss the constraints on $\tb$ within a so-called 
maximal mixing ``benchmark'' scenario~\cite{lep2CarZer}. In this scenario,
$\mt$ was kept fixed at the value $\mt = 175 \gev$, a large value of 
$\msusy$ was chosen, $\msusy = 1 \tev$, and the mixing parameter in the 
stop sector was fixed 
in order to maximize the stop-induced radiative corrections
to the lightest $\cp$-even Higgs-boson mass as a function of $\tan\beta$, 
$\mh(\tb)$. For a
recent analysis within this framework, see e.g.\ \citere{Higgsgroup}.

In this note we shall define an improved version of the maximal mixing
benchmark scenario that keeps many of the features of the previous one,
but maximizes also the chargino and neutralino contributions by taking
small values of the $|\mu|$ and $M_2$ parameters, while yielding
chargino masses which are beyond the reach of LEP2. 
This scenario maximizes the Higgs-boson mass as a function of
$\tb$ for fixed $\mt$ and $\msusy$ ($\mhmax$ scenario), and
should therefore be useful in order to derive conservative bounds on $\tb$.
The $\mhmax$ scenario defined here is close to the one recently proposed 
in \citere{tbexcl}, where it was analyzed how the previous benchmark
scenario~\cite{Higgsgroup}
should be modified in order to incorporate the maximal values
of $\mh(\tb)$. An analysis of the experimental lower bound on $\tb$,
studying its dependence on the $\Stop$-mass eigenvalues and the mixing 
angle  was performed in \citere{ccpw}. 
The values of $\mu$ and $M_2$ in \citere{ccpw} were similar to those
proposed here for the $\mhmax$ scenario.

In the following we will consider the $\mhmax$ scenario as well as the 
corresponding scenario with
vanishing mixing in the scalar top sector. We furthermore suggest a
third scenario, in which a relatively large value of $|\mu|$ is adopted,
leading to interesting phenomenological consequences.

In all benchmark scenarios we fix the top-quark mass to its 
experimental central value, 
\BE
\mt = \mtexp~  ( = 174.3 \gev ) ,
\label{mtexp}
\EE
where we have indicated the current value for completeness.
It should be kept in mind that internally the codes \subh\ and \fh\ make
use of the running top-quark mass, $\mtbar$. In comparing results of
different codes it is essential that not only the input value for the
top-quark pole mass is the same, but also the relation employed for
deriving the running top-quark mass. In \subh\ and \fh\ $\mtbar$ is
calculated from $\mt$ according to \refeq{mtrun}, taking into account
corrections up to $\oas$.

Although the soft SUSY breaking parameter $\msusy$ is 
renormalization-scheme-dependent, the numerical effect of the scheme 
dependence is rather small in general. We have checked that for the
scenarios below the numerical difference of the corresponding values
of the parameter $\msusy$ in the two schemes
lies within about 4\%. We therefore do not distinguish between the 
parameters in the two schemes and define
\BE
\msusy^{\MS} \approx \msusy^{\OS} =: \msusy .
\label{msusy}
\EE

%%%%%%%%%%%%%%%%%%%%%%%%%%%%%%%%%%%%%%%%%%%%%%%%%%%%%%%%%%%%%%
%%%%%%%%%%%%%%%%%%%%%%%%%%%%%%%%%%%%%%%%%%%%%%%%%%%%%%%%%%%%%%

\subsection{The $\mhmax$ scenario}

In this benchmark scenario the parameters are chosen such that the
maximum possible Higgs-boson mass as a function of $\tb$ is obtained
(for fixed $\msusy$, $\mt$ given by its experimental central value, 
and $\MA$ set to its
maximal value in this scenario, $\MA = 1 \tev$). 
The parameters are:
\BEA
\msusy &=& 1 \tev \non \\
\mu &=& -200 \gev \non \\
M_2 &=& 200 \gev \non \\
\label{mhmax}
\mgl &=& 0.8\,\msusy  \\
\MA &\leq& 1000 \gev \non \\
\Xt^{\OS} &=& 2\, \msusy  \quad \mbox{(FD calculation)} \non \\
\Xt^{\MS} &=& \sqrt{6}\, \msusy \quad \mbox{(RG calculation)} \non \\
\Ab &=& \At~. \non
\EEA

%The contour in the $\mh$--$\tb$-plane for $\MA = 1 \tev$ corresponds to
%the boundary of the region that is ``theoretically not allowed'', since
%in this region the $\mh(\tb)$ values are outside the range predicted
%by the MSSM.

The values for $\Xt$ in the FD calculation (\fh) and in the RG
calculation (\subh) specify the mixing in the scalar top sector in both
approaches in such a way that $\mh$ becomes maximal. 
The values of $\mu$ and $M_2$ are close to their experimental lower
bounds. 
%the value of $\mu$ is chosen such that it will not be ruled out by LEP2. 
Slightly higher Higgs-boson masses are obtained for smaller 
$|\mu|$ and smaller $M_2$. The sign of $\mu$ has only a  small effect 
in this scenario. 

One should take into account that the maximal value of the lightest 
$\cp$-even Higgs-boson mass would increase with respect to the $\mhmax$
benchmark scenario if, for instance, the $1 \sigma$ upper bound on the 
experimental value of the top-quark mass were considered,
or if the third generation squark masses were larger than the ones chosen
in the benchmark scenario.

%%%%%%%%%%%%%%%%%%%%%%%%%%%%%%%%%%%%%%%%%%%%%%%%%%%%%%%%%%%%%%
%%%%%%%%%%%%%%%%%%%%%%%%%%%%%%%%%%%%%%%%%%%%%%%%%%%%%%%%%%%%%%

\subsection{The no-mixing scenario}

This benchmark scenario is the same as the $\mhmax$ scenario, but with
vanishing mixing in the $\Stop$-sector. The parameters are:
\BEA
\msusy &=& 1 \tev \non \\
\mu &=& -200 \gev \non \\
M_2 &=& 200 \gev \non \\
\label{nomix}
\mgl &=& 0.8\,\msusy \\
\MA &\leq& 1000 \gev \non \\
\Xt^{\OS} &=& 0  \quad \mbox{(FD calculation)} \non \\
\Xt^{\MS} &=& 0 \quad \mbox{(RG calculation)} \non \\
\Ab &=& \At~, \non
\EEA
where we have neglected the difference between $\Xt^{\OS}$ and
$\Xt^{\MS}$, which is of minor importance in this scenario.

The difference of the $\mh$ values in the $\mhmax$ and the no-mixing
scenario is purely an effect of the mixing in the scalar top sector.
For a common $\msusy$ and low values of $|\mu|$ and $M_2$, as assumed
above, restrictions on the mixing parameters in the $\Stop$-sector as a
function of $\tb$ can be derived by demanding the Higgs-boson mass to be
above the experimental limit. This is due to the fact that for low 
values of $\tb$ experimentally acceptable values of $\mh$ can only be 
achieved for non-vanishing mixing parameters in the scalar top sector.

%By comparing the no-mixing scenario with the search limits on $\mh$
%restrictions on the mixing in the $\Stop$-sector, depending on $\tb$,
%can be derived.

%%%%%%%%%%%%%%%%%%%%%%%%%%%%%%%%%%%%%%%%%%%%%%%%%%%%%%%%%%%%%%
%%%%%%%%%%%%%%%%%%%%%%%%%%%%%%%%%%%%%%%%%%%%%%%%%%%%%%%%%%%%%%

\subsection{The large $\mu$ scenario}

This benchmark scenario is characterized by a relatively large value of
$|\mu|$ (compared to $\msusy$). We furthermore adopt a relatively small
value of $\msusy$ and moderate mixing in the scalar top sector.
The parameters are:
\BEA
\msusy &=& 400 \gev \non \\
%\mu &=& 2\,\msusy \non \\
\mu &=& 1 \tev \non \\
M_2 &=& 400 \gev \non \\
\label{lowtb}
\mgl &=& 200 \gev \\
\MA &\leq& 400 \gev \non \\
% \tb &\leq& 40  \non \\
\Xt^{\OS} &=& -300 \gev  \quad \mbox{(FD calculation)} \non \\
\Xt^{\MS} &=& -300 \gev \quad \mbox{(RG calculation)} \non \\
\Ab &=& \At~ \non \\
\mb &\equiv& \mb(\mt) = 3 \gev \quad \mbox{(FD calculation)}. \non
\EEA
Here we have neglected the difference between $\Xt^{\OS}$ and 
$\Xt^{\MS}$. This will slightly affect the comparison between the FD and
the RG result, but will be of minor relevance for the general features
of the large $\mu$ scenario which are discussed in the following.
The value of the bottom mass, $\mb(\mt) = 3 \gev$, specified for the 
FD calculation is chosen in order to absorb higher-order QCD corrections
that are important to keep the effects of large mixing in the scalar bottom 
sector, which occur for large $\mu$ and $\tb$, under control. In \subh\
this is already taken into account internally.
%The
%upper bound on the value of the ratio of vacuum expectation values,
%$\tan\beta$, comes from the requirement of getting acceptable values
%for the precision electroweak observables. 

As a consequence of the relatively low values of $\msusy$ and the mixing
parameter in the $\Stop$-sector chosen in this scenario, considerably lower
Higgs-boson masses are obtained compared to the $\mhmax$ scenario. 
Therefore, in the large $\mu$ scenario defined here LEP2 has the potential 
of covering the whole $\mh$--$\tb$-plane. It should furthermore be noted 
that for large values of $\tb$ in this scenario radiative corrections from 
the scalar bottom sector become important, which, for instance, 
lead to a decrease of the predicted value for $\mh$ for moderate or
large values of the CP-odd Higgs mass $M_A \gsim 150$ GeV.

%Compared to the $\mhmax$ scenario, as a consequence of the choices for
%$\msusy$ and the mixing in the $\Stop$-sector, LEP2 can access a
%considerably larger region of the $\mh$--$\tb$-plane within the large
%$\mu$ scenario. Furthermore, for large values of $\tb$ and $\mu$ radiative
%corrections from the scalar bottom sector become important, which lead
%to a decrease of the predicted value for $\mh$.

On the other hand, this scenario also gives rise to regions in the MSSM
parameter space where the Higgs-boson detection might be difficult,
since there exist ``pathological'' points for which either 
$BR(\hbb) \to 0$ or $BR(\Hbb) \to 0$~\cite{wells,htobb0}. 
Although the relevant Higgs-boson
mass will in principle be within the kinematically accessible region, the
non-standard decay signatures may lead to difficulties in actually
detecting the particle. For a recent analysis in this context see 
\citeres{htobb0,comp}.

The condition, whether corrections in the Higgs sector lead to a
vanishing effective coupling $h b \bar b$ or $H b \bar b$ (and
consequently to $BR(\hbb) \to 0$ or $BR(\Hbb) \to 0$), depends in
particular on the signs and magnitudes of $(\mu A_t)$ and $(\mu A_b)$ 
and also on the value of $|A_t|$~\cite{htobb0}.
Changing the sign of $X_t$  in \refeq{lowtb} leads to a scenario 
with similar $\mh$~values, where the radiative corrections in the 
Higgs sector do not
give rise to ``pathological'' points in the $\mh$--$\tb$-plane with
vanishing branching ratio of the $\cp$-even Higgs-boson decays into
bottom quarks.

For large $\mu$, $\tb$, and $\mgl$ large SUSY-QCD corrections to the 
$hb\bar b$ vertex are possible that could make a perturbative
calculation questionable~\cite{deltamb,babu,hff1l,htobb0,comp}. 
Even for the relatively 
low value of $\mgl = 200$~GeV chosen in the present scenario, very large 
vertex corrections from gluino exchange appear in the large $\tb$
region, which can also lead to a suppression of the branching ratio into
bottom quarks. 
It should furthermore be noted that not only the SUSY-QCD vertex 
corrections but also the genuine electroweak vertex contributions
can become relevant. In order to obtain reliable predictions in this
region of parameters, 
the inclusion of leading higher-order contributions is important.
A proper treatment of the vertex corrections in the region of
large values of $\tb$   
will be incorporated in the versions of the programs \fhtwo\ and \subhtwo\
available for public use in the near future.

The same kind of SUSY-QCD corrections affects the value of the
electroweak precision parameter
$\Delta \rho$, which, for the values of the parameters chosen in
this scenario, can exceed the experimentally allowed values
for extra SUSY contributions, $\Delta \rho^{\rm SUSY} \lsim 10^{-3}$,
for very large values of $\tb$. 
The value of $\Delta \rho$, based on a two-loop calculation~\cite{drho}, is
given as an output of \fh\ as a consistency check of the calculation.
A thorough treatment of the higher order SUSY-QCD corrections is 
also important in this case.

Besides the suppression of the main decay channel, a problem for
detecting the MSSM Higgs bosons can of course also arise from a
suppression of the kinematically favored production cross section,
i.e.\ $e^+e^- \to hZ$ or $e^+e^- \to hA$~\cite{exper}. 
For instance, at LEP2
this behaviour occurs in the $\mhmax$ scenario
for relatively large values 
of $\tb$ and values of $M_A$ 
such that the lightest $\cp$-even Higgs-boson mass is just above the
kinematical  threshold of the $hA$ channel. In this region of
parameters, the  lightest CP-even 
Higgs boson  is within the kinematical reach of the $hZ$ channel, but 
its coupling to the $Z$~boson is suppressed.
For the large $\mu$ scenario, instead, we find that at least one
of the production channels $e^+e^- \to hZ, \, HZ, \, hA$ should
always be open within the kinematical reach 
with a sufficiently high rate. The reason for this can
qualitatively be understood
from the fact that the cross sections for the above 
production channels are approximately proportional to 
$\sin^2(\beta-\alpha)$, $\cos^2(\beta-\alpha)$, $\cos^2(\beta-\alpha)$,
respectively, and from the relation~\cite{comp}
\begin{equation}
\left.
m_h^2 \sin^2(\beta-\alpha) + m_H^2 \cos^2(\beta-\alpha) = 
m_h^2 \right|_{M_A^2 \gg M_Z^2}.
\end{equation}
In the above, the quantities on the left-hand side are given as
functions of arbitrary values of $\MA$ and the other MSSM
parameters, while the right-hand side is the square of the 
lightest $\cp$-even Higgs-boson mass for $\MA^2 \gg \MZ^2$ and the same
values of the other MSSM parameters, i.e.\ the upper bound on $\mh$ 
for this set of parameters. 
In the large $\mu$ scenario, the upper bound on $\mh$
is about 107 GeV, which is within the kinematical reach of LEP2, 
and is only obtained for relatively large values of $\tan\beta$. 
Therefore, the suppression of the $hZ$ or $HZ$
production cross section by very small values
of one of these mixing angles implies that the complementary cross
section will be of the order of the Standard Model one and that
the corresponding Higgs boson is within the LEP2 kinematical reach.

%%%%%%%%%%%%%%%%%%%%%%%%%%%%%%%%%%%%%%%%%%%%%%%%%%%%%%%%%%%%%%
%%%%%%%%%%%%%%%%%%%%%%%%%%%%%%%%%%%%%%%%%%%%%%%%%%%%%%%%%%%%%%

\section{Conclusions}

We have suggested three benchmark scenarios for the Higgs-boson 
search at LEP2, which improve and extend the previous benchmark
definitions used in the literature.
The $\mhmax$ scenario yields the theoretical upper bound of $\mh$ in
the MSSM as a function of $\tb$ for fixed $\mt$ and $\msusy$. It thus
allows to derive conservative constraints on $\tb$ from the Higgs-boson
search under the assumption that $\mt$ is given by its
experimental central value and $\msusy = 1$ TeV. 
In the no-mixing scenario the mixing in the scalar top sector is
chosen to be zero, while the other parameters are the same as in the 
$\mhmax$ scenario. Comparing the two scenarios allows to investigate the
effects of mixing in the scalar top sector. As a new benchmark scenario,
we propose a scenario which is characterized by a relatively large value
of $|\mu|$. Moderate values are chosen for $\msusy$ and the mixing parameter
in the $\Stop$-sector. The values of the 
Higgs-boson masses obtained in this scenario are
such that in principle a complete coverage of the $\mh$--$\tb$-plane
would be possible at LEP2. However, the scenario contains parameter
regions in which the $BR(\hbb)$ or $BR(\Hbb)$
is suppressed and which therefore 
will be difficult to access, although the corresponding Higgs-boson mass 
would be within the kinematical reach. Thus, other decay modes of the
Higgs boson beyond the $b \bar b$ channel should carefully be considered in
this region of parameters.

\section*{Acknowledgements}
We would like to thank H.\ Haber, W.\ Hollik, S.\ Mrenna, A.~Pilaftsis, 
S.~Pokorski, M.~Quiros and the members of the LEP Higgs working group 
for very interesting comments and discussions.
S.H.\ also thanks the OPAL group for financial support during his visit at
CERN, where part of this work was done.

%%%%%%%%%%%%%%%%%%%%%%%%%%%%%%%%%%%%%%%%%%%%%%%%%%%%%%%%%%%%%%
%%%%%%%%%%%%%%%%%%%%%%%%%%%%%%%%%%%%%%%%%%%%%%%%%%%%%%%%%%%%%%

%\newpage

%%%%%%%%%%%%%%%%%%%%%%%%%%%%%%%%%%%%%%%%%%%%%%%%%%%%%%%%%%%%%%
%%%%%%%%%%%%%%%%%%%%%%%%%%%%%%%%%%%%%%%%%%%%%%%%%%%%%%%%%%%%%%


\begin{thebibliography}{00}  



\bibitem{mhiggs1l} H.~Haber and R.~Hempfling,
                   {\em Phys. Rev. Lett.} {\bf 66} (1991) 1815;\\
                   Y.~Okada, M.~Yamaguchi and T.~Yanagida,
                   {\em Prog. Theor. Phys.} {\bf 85} (1991) 1;\\
                   J.~Ellis, G.~Ridolfi and F.~Zwirner,
                   {\em Phys. Lett.} {\bf B 257} (1991) 83; 
                   {\em Phys. Lett.} {\bf B 262} (1991) 477;\\
                   R.~Barbieri and M.~Frigeni,
                   {\em Phys. Lett.} {\bf B 258} (1991) 395.

\bibitem{mhiggsf1l} P.~Chankowski, S.~Pokorski and J.~Rosiek,
                    {\em Nucl. Phys.} {\bf B 423} (1994) 437;\\
                    A.~Dabelstein, 
                    {\em Nucl. Phys.} {\bf B 456} (1995) 25,
                    hep-ph/9503443;
                    {\em Z. Phys.} {\bf C 67} (1995) 495,
                    hep-ph/9409375;\\
                    J.~Bagger, K.~Matchev, D.~Pierce and R.~Zhang,
                    {\em Nucl. Phys.} {\bf B 491} (1997) 3,
                    hep-ph/9606211.


\bibitem{LL}
R.~Barbieri, M.~Frigeni and F.~Caravaglios, \PLB{258}{91}{167};\\
Y.~Okada, M.~Yamaguchi and T.~Yanagida, \PLB{262}{91}{54};\\
% \bibitem{CSW} 
M.~Carena, K.~Sasaki and C.E.M.~Wagner, \NPB{381}{92}{66};\\
% \bibitem{CPR} 
P.~Chankowski, S.~Pokorski and J.~Rosiek, \PLB{281}{92}{100};\\
% \bibitem{HH}
H.E.~Haber and R.~Hempfling, \PRD{48}{93}{4280};\\
%\bibitem{Kelley} 
S.~Kelley, J.~Lopez, D.~Nanopoulos, H.~Pois and K.~Yuan, 
\NPB{398}{93}{3};\\
% \bibitem{LP} 
P.~Langacker and N.~Polonsky, \PRD{50}{94}{2199};\\
% \bibitem{KYS}
J.~Kodaira, Y.~Yasui and K.~Sasaki, \PRD{50}{94}{7035};\\
%\bibitem{mhiggsRG1}  
J.~Casas, J.~Espinosa, M.~Quir\'os and A.~Riotto,
                     {\em Nucl. Phys.} {\bf B 436} (1995) 3,
                     E: {\em ibid.} {\bf B 439} (1995) 466,
                     hep-ph/9407389;\\
%
A. Pilaftsis and C.E.M. Wagner, {\em Nucl. Phys.} {\bf B 553} (1999) 3.


\bibitem{mhiggsRG1a}  M.~Carena, J.~Espinosa, M.~Quir\'os and C.E.M.~Wagner,
                      {\em Phys. Lett.} {\bf B 355} (1995) 209,
                      hep-ph/9504316.
                      


\bibitem{mhiggsRG1b}  M.~Carena, M.~Quir\'os and C.E.M.~Wagner,
                      {\em Nucl. Phys.} {\bf B 461} (1996) 407,
                      hep-ph/9508343.

\bibitem{mhiggsRG2} H.~Haber, R.~Hempfling and A.~Hoang,
                    {\em Z. Phys.} {\bf C 75} (1997) 539,
                    hep-ph/9609331.

\bibitem{hoanghempfling} R.~Hempfling and A.~Hoang,
                         {\em Phys. Lett.} {\bf B 331} (1994) 99,
                         hep-ph/9401219.

\bibitem{zhang} R.-J.~Zhang, 
                {\em Phys. Lett.} {\bf B 447} (1999) 89,
                hep-ph/9808299.

\bibitem{mhiggsletter} S.~Heinemeyer, W.~Hollik and G.~Weiglein,
                    {\em Phys. Rev.} {\bf D 58} (1998) 091701,
                    hep-ph/9803277;
                    {\em Phys. Lett.} {\bf B 440} (1998) 296,
                    hep-ph/9807423.

\bibitem{mhiggslong} S.~Heinemeyer, W.~Hollik and G.~Weiglein,
                     {\em Eur. Phys. Jour.} {\bf C 9} (1999) 343, 
                     DOI~10.1007/s100529900006,
                     hep-ph/9812472.


\bibitem{feynhiggs} S.~Heinemeyer, W.~Hollik and G.~Weiglein,
                    to appear in {\em Comp. Phys. Comm.}, 
                    hep-ph/9812320.

\bibitem{bse} M.~Carena, H.~Haber, S.~Heinemeyer, W.~Hollik, C.E.M.~Wagner
              and G.~Weiglein,
              {\em in preparation}.

\bibitem{mhiggslle} S.~Heinemeyer, W.~Hollik and G.~Weiglein,
                    {\em Phys. Lett.} {\bf B 455} (1999) 179,
                    hep-ph/9903404.

\bibitem{lep2CarZer} 
M.~Carena, P.M.~Zerwas et al., in {\em Physics at LEP2},
eds.\ G.~Altarelli, T.~Sj\"ostrand and F.~Zwirner, CERN 96-01,
p.~351.

\bibitem{Higgsgroup} The LEP working group for Higgs boson searches,
                     CERN-EP/99-060.

\bibitem{tbexcl} S.~Heinemeyer, W.~Hollik and G.~Weiglein, 
                 DESY 99-120, hep-ph/9909540.

\bibitem{ccpw} M.~Carena, P.~Chankowski, S.~Pokorski and C.E.M.~Wagner,
               {\em Phys. Lett.} {\bf B 441} (1998) 205, hep-ph/9805349.




\bibitem{wells}
 W.~Loinaz and J.D.~Wells, {\em Phys. Lett.} {\bf B445} (1998) 178; \\
H.~Baer and J.D.~Wells, {\em Phys.Rev.} {\bf D57} (1998) 4446.

\bibitem{htobb0} M.~Carena, S.~Mrenna and C.E.M.~Wagner, 
                 {\em Phys. Rev.} {\bf D 60} (1999) 075010, hep-ph/9808312.


\bibitem{comp}   M.~Carena, S.~Mrenna and C.E.M.~Wagner, 
                 CERN-TH/99-203, hep-ph/9907422.

\bibitem{deltamb} L.~Hall, R.~Rattazzi and U.~Sarid,  {\em Phys. Rev.}
{\bf D 50} (1994) 7048; \\
R.~Hempfling, {\em Phys. Rev.} {\bf D 49} (1994) 6168; \\
M.~Carena, M.~Olechowski, S.~Pokorski and C.E.M.~Wagner,
{\em Nucl. Phys.} {\bf B 426} (1994) 269.  


\bibitem{babu}
K.S.~Babu and C.~Kolda, {\em Phys. Lett.} {\bf B 451} (1999) 77; \\
F.~Borzumati, G.R.~Farrar, N.~Polonsky and S.~Thomas, 
{\em Nucl. Phys.} {\bf B 555} (1999) 53. 


\bibitem{hff1l} A.~Dabelstein,
                {\em Nucl. Phys.} {\bf B 456} (1995) 25,
                hep-ph/9503443;\\
                J.A.~Coarasa, R.A.~Jim\'enez and J.~Sol\`a,
                {\em Phys. Lett.} {\bf B 389} (1996) 312,
                hep-ph/9511402;
                S.~Heinemeyer, W.~Hollik and G.~Weiglein, KA-TP-9-1999.
                %{\em in preparation}.

\bibitem{drho}     A.~Djouadi, P.~Gambino, S.~Heinemeyer, W.~Hollik,
                    C.~J\"unger and G.~Weiglein,
                    {\em Phys. Rev. Lett.} {\bf 78} (1997) 3626,
                    hep-ph/9612363;
                    {\em Phys. Rev.} {\bf D 57} (1998) 4179,
                    hep-ph/9710438.


\bibitem{exper}
J.~Rosiek and A.~Sopczak, {\em Phys. Lett.} {\bf B 341} (1995) 419;\\
A.~Sopczak, {\em Eur. Phys.} {\bf C9} (1999) 107;\\
ALEPH collaboration, {\em Phys. Lett.} {\bf B 440} (1998) 419.

\end{thebibliography}
\end{document}